\documentclass[12pt]{article}
\usepackage{psfig}
\usepackage{times}

\makeatletter

\def\revtex@ver{1.6}		
\def\revtex@date{12 Aug 93}	
\def\revtex@org{PASP}		
\def\revtex@jnl{}		
\def\revtex@genre{conference proceedings}	
\def\revtex@pageid{\xdef\@thefnmark{\null}
\@footnotetext{This \revtex@genre\space was prepared with the
\revtex@org\space \revtex@jnl\space Rev\TeX\ macros v\revtex@ver.}}
\ifnum\@ptsize<1
\fi
\def\ps@paspcstitle{\let\@mkboth\@gobbletwo
\def\@oddhead{\null{\footnotesize\it\@slug}\hfil}
\def\@oddfoot{\rm\hfil\thepage\hfil}
\let\@evenhead\@oddhead\let\@evenfoot\@oddfoot
}
\def\ps@myheadings{\let\@mkboth\@gobbletwo
\def\@oddhead{\hbox{}\hfil\sl\rightmark\hskip 1in\rm\thepage}%
\def\@oddfoot{}%
\def\@evenhead{\rm\thepage\hskip 1in\sl\leftmark\hfil\hbox{}}%
\def\@evenfoot{}\def\sectionmark##1{}\def\subsectionmark##1{}}
\def\@leftmark#1#2{\sec@upcase{#1}}
\def\@rightmark#1#2{\sec@upcase{#2}}
\def\@singleleading{0.9}
\def\@doubleleading{1.6}
\def\baselinestretch{\@singleleading}
\def\tightenlines{\def\baselinestretch{\@singleleading}}
\def\loosenlines{\def\baselinestretch{\@doubleleading}}
\clubpenalty\@M
\widowpenalty\@M
\def\@journalname{ASP Conference Series}
\def\cpr@holder{Astronomical Society of the Pacific}
\def\@jourvol{10000}
\def\cpr@year{1997}
\def\vol@title{Cool Stars, Stellar Systems and the Sun}
\def\vol@author{J.\ A.\ Bookbinder and R.\ A.\ Donahue, eds.}
\let\journalid=\@gobbletwo
\let\articleid=\@gobbletwo
\let\received=\@gobble
\let\accepted=\@gobble
\def\@slug{{\tabcolsep\z@\begin{tabular}[t]{l}\vol@title\\
\@journalname, Vol.\ \@jourvol, \cpr@year\\
\vol@author
\end{tabular}}
}
\def\paspconf@frontindent{.45in}
\def\title#1{\vspace*{1.0\baselineskip}
\@tempdima\textwidth \advance\@tempdima by-\paspconf@frontindent
\hfill
\parbox{\@tempdima}
	{\pretolerance=10000\raggedright\large\bf\sec@upcase{#1}}\par
\vspace*{1\baselineskip}\thispagestyle{title}}
\def\author#1{\vspace*{1\baselineskip}
\@tempdima\textwidth \advance\@tempdima by-\paspconf@frontindent
\hfill
\parbox{\@tempdima}
{\pretolerance=10000\raggedright{#1}}\par}
\def\affil#1{\vspace*{.5\baselineskip}
\@tempdima\textwidth \advance\@tempdima by-\paspconf@frontindent
\hfill
\parbox{\@tempdima}
{\pretolerance=10000\raggedright{\it #1}}\par}

\def\abstract{\vspace*{1.3\baselineskip}\bgroup\leftskip\paspconf@frontindent
\noindent{\bf\sec@upcase{Abstract.}}\hskip 1em}
\def\endabstract{\par\egroup\vspace*{1.4\baselineskip}}
\skip\footins 4ex plus 1ex minus .5ex
\long\def\@makefntext#1{\noindent\hbox to\z@{\hss$^{\@thefnmark}$}#1}
\def\tablenotemark#1{\rlap{$^{#1}$}}
\def\tablenotetext#1#2{
\@temptokena={\vspace{.5ex}{\noindent\llap{$^{#1}$}#2}\par}
\@temptokenb=\expandafter{\tblnote@list}
\xdef\tblnote@list{\the\@temptokenb\the\@temptokena}}
\def\spewtablenotes{
\ifx\tblnote@list\@empty
\else
\let\@temptokena=\tblnote@list
\gdef\tblnote@list{\@empty}
\vspace{4.5ex}
\footnoterule
\vspace{.5ex}
{\footnotesize\@temptokena}
\fi}
\newtoks\@temptokenb
\def\tblnote@list{}
\def\endtable{\spewtablenotes\end@float}
\@namedef{endtable*}{\spewtablenotes\end@dblfloat}

\def\thefigure{\@arabic\c@figure}
\def\fnum@figure{Figure \thefigure.}
\def\thetable{\@arabic\c@table}
\def\fnum@table{Table \thetable.}
\long\def\@makecaption#1#2{
\vskip 10pt
\setbox\@tempboxa\hbox{#1\hskip 1.5em #2}
\let\@tempdima=\hsize \advance\@tempdima by -2em
\ifdim \wd\@tempboxa >\@tempdima
	{\leftskip 2em
	#1\hskip 1.5em #2\par}
\else
	\hbox to\hsize{\hskip 2em\box\@tempboxa\hfil}
\fi}
\def\fps@figure{tbp}
\def\fps@table{htbp}
\let\keywords=\@gobble
\let\subjectheadings=\@gobble
\def\upper{\def\sec@upcase##1{\uppercase{##1}}}
\def\sec@upcase#1{\relax#1}
\def\section{\@startsection {section}{1}{\z@}{-4.2ex plus -1ex minus
-.2ex}{2.2ex plus .2ex}{\normalsize\bf}}
\def\subsection{\@startsection{subsection}{2}{\z@}{-2.2ex plus -1ex minus
-.2ex}{1.1ex plus .2ex}{\normalsize\bf}}

\def\subsubsection{\@startsection{subsubsection}{3}{\z@}{-2.2ex plus
-1ex minus -.2ex}{-1.2em}{\normalsize\it}}
\def\thesection{\@arabic\c@section.}
\def\thesubsection{\thesection\@arabic\c@subsection.}
\def\thesubsubsection{\thesubsection\@arabic\c@subsubsection.}
\def\@sect#1#2#3#4#5#6[#7]#8{\ifnum #2>\c@secnumdepth
\def\@svsec{}\else
\refstepcounter{#1}\edef\@svsec{\csname the#1\endcsname\hskip 1em }\fi
\@tempskipa #5\relax
\ifdim \@tempskipa>\z@
\begingroup #6\relax
\@hangfrom{\hskip #3\relax\@svsec}{\interlinepenalty \@M \sec@upcase{#8}\par}%
\endgroup
\csname #1mark\endcsname{#7}\addcontentsline
{toc}{#1}{\ifnum #2>\c@secnumdepth \else
\protect\numberline{\csname the#1\endcsname}\fi
#7}\else
\def\@svsechd{#6\hskip #3\@svsec #8\csname #1mark\endcsname
{#7}\addcontentsline
{toc}{#1}{\ifnum #2>\c@secnumdepth \else
\protect\numberline{\csname the#1\endcsname}\fi
#7}}\fi
\@xsect{#5}}
\def\@ssect#1#2#3#4#5{\@tempskipa #3\relax
\ifdim \@tempskipa>\z@
\begingroup #4\@hangfrom{\hskip #1}{\interlinepenalty \@M \sec@upcase{#5}\par}\endgroup
\else \def\@svsechd{#4\hskip #1\relax #5}\fi
\@xsect{#3}}
\def\acknowledgments{\@startsection{paragraph}{4}{1em}
{1ex plus .5ex minus .5ex}{-1em}{\bf}{\sec@upcase{Acknowledgments.}}}

\def\qanda@heading{Discussion}
\newif\if@firstquestion \@firstquestiontrue

\def\mathwithsecnums{
\@newctr{equation}[section]
\def\theequation{\hbox{\normalsize\arabic{section}-\arabic{equation}}}}
\def\references{\section*{References}
\bgroup\parindent=0pt\parskip=.5ex
\def\refpar{\par\hangindent=3em\hangafter=1}}
\def\endreferences{\refpar\egroup}
\def\thebibliography{\section*{References}
\list{\null}{\leftmargin 3em\labelwidth 0pt\labelsep 0pt\itemindent -3em
\usecounter{enumi}}
\def\refpar{\relax}
\def\newblock{\hskip .11em plus .33em minus .07em}
\sloppy\clubpenalty4000\widowpenalty4000
\sfcode`\.=1000\relax}

\def\@biblabel#1{\relax}
\def\@cite#1#2{#1\if@tempswa , #2\fi}

\def\@citex[#1]#2{\if@filesw\immediate\write\@auxout{\string\citation{#2}}\fi
\def\@citea{}\@cite{\@for\@citeb:=#2\do
{\@citea\def\@citea{,\penalty\@m\ }\@ifundefined
{b@\@citeb}{\@warning
{Citation `\@citeb' on page \thepage \space undefined}}%
{\csname b@\@citeb\endcsname}}}{#1}}
\let\jnl@style=\rm
\def\ref@jnl#1{{\jnl@style#1\/}}
\def\acta{\ref@jnl{Acta~Astron.}}
\def\aj{\ref@jnl{AJ}}			
\def\araa{\ref@jnl{ARA\&A}}		
\def\apj{\ref@jnl{ApJ}}			
\def\apjl{\ref@jnl{ApJ}}		
\def\apjs{\ref@jnl{ApJS}}		
\def\ao{\ref@jnl{Appl.Optics}}		
\def\apss{\ref@jnl{Ap\&SS}}		
\def\aap{\ref@jnl{A\&A}}		
\def\aapr{\ref@jnl{A\&A~Rev.}}		
\def\aaps{\ref@jnl{A\&AS}}		
\def\azh{\ref@jnl{AZh}}			
\def\baas{\ref@jnl{BAAS}}		
\def\csw6{\ref@jnl{in ASP Conf.~Ser. 9, The Sixth Cambridge Workshop on 
	Cool Stars, Stellar Systems and the Sun, ed. G. Wallerstein 
	(San Francisco: ASP)}}
\def\csw7{\ref@jnl{in ASP Conf.~Ser. 26, The Seventh Cambridge Workshop on 
	Cool Stars, Stellar Systems and the Sun, eds. M.S. Giampapa \& 
	J.A. Bookbinder (San Francisco: ASP)}}
\def\csw8{\ref@jnl{in ASP Conf.~Ser. 64, The Eighth Cambridge Workshop on 
	Cool Stars, Stellar Systems and the Sun, ed. J.-P. Caillault
	(San Francisco: ASP)}}
\def\csw9{\ref@jnl{in ASP Conf.~Ser. 109, The Ninth Cambridge Workshop on
	Cool Stars, Stellar Systems and the Sun, eds. R. Pallavicini \&
	A.K. Dupree (San Francisco: ASP)}}
\def\csw10{\ref@jnl{in ASP Conf.~Ser. ??, The Tenth Cambridge Workshop on
	Cool Stars, Stellar Systems and the Sun, eds. J.A. Bookbinder \&
	R.A. Donahue (San Francisco: ASP)}}
\def\jrasc{\ref@jnl{JRASC}}		
\def\memras{\ref@jnl{MmRAS}}		
\def\mnras{\ref@jnl{MNRAS}}		
\def\pra{\ref@jnl{Phys.Rev.A}}		
\def\prb{\ref@jnl{Phys.Rev.B}}		
\def\prc{\ref@jnl{Phys.Rev.C}}		
\def\prd{\ref@jnl{Phys.Rev.D}}		
\def\prl{\ref@jnl{Phys.Rev.Lett}}	
\def\pasp{\ref@jnl{PASP}}		
\def\pasj{\ref@jnl{PASJ}}		
\def\qjras{\ref@jnl{QJRAS}}		
\def\skytel{\ref@jnl{S\&T}}		
\def\solphys{\ref@jnl{Solar~Phys.}}	
\def\sovast{\ref@jnl{Soviet~Ast.}}	
\def\ssr{\ref@jnl{Space~Sci.Rev.}}	
\def\zap{\ref@jnl{ZAp}}

\def\la{\mathrel{\hbox{\rlap{\hbox{\lower4pt\hbox{$\sim$}}}\hbox{$<$}}}}
\def\ga{\mathrel{\hbox{\rlap{\hbox{\lower4pt\hbox{$\sim$}}}\hbox{$>$}}}}

\def\ion#1#2{#1$\;${\small\rm\expandafter\uppercase\expandafter{\romannumeral #2}\relax}}

\newcount\lecurrentfam
\def\LaTeX{\lecurrentfam=\the\fam \leavevmode L\raise.42ex
\hbox{$\fam\lecurrentfam\scriptstyle\kern-.3em A$}\kern-.15em\TeX}

\newbox\pt@box
\newdimen\pt@width
\newcount\pt@line
\newcount\pt@column
\newcount\pt@nlines
\newcount\pt@ncol
\newcount\pt@page
\def\colhead#1{\multicolumn{1}{c}{#1}\pt@addcol}
\def\tablecolumns#1{\pt@column=#1\pt@ncol=#1\gdef\pt@addcol{\relax}}
\def\tablecaption#1{\gdef\pt@caption{#1}} \def\pt@caption{\relax}
\def\tablehead#1{\gdef\pt@head{\hline\hline\relax\\[-1.7ex]
#1\hskip\tabcolsep\\[.7ex]\hline\relax\\[-1.5ex]}} \def\pt@head{\relax}
\def\tabletail#1{\gdef\pt@tail{#1}} \def\pt@tail{\relax}
\def\tablewidth#1{\pt@width=#1} \pt@width\textwidth
\def\tableheadfrac#1{\gdef\pt@headfrac{#1}} \def\pt@headfrac{.1}

\def\pt@calcnlines{\@tempdima\pt@headfrac\textheight
\@tempdimb\textheight\advance\@tempdimb by-\@tempdima
\@tempdima\arraystretch\baselineskip
\divide\@tempdimb by\@tempdima
\global\pt@nlines\@tempdimb}

\def\pt@tabular{\hbox \bgroup $\let\@acol\@ptabacol
\let\@classz\@tabclassz
\let\@classiv\@tabclassiv \let\\\@tabularcr\@tabarray}

\def\@ptabacol{\edef\@preamble{\@preamble \hskip \tabcolsep\tabskip\fill}}

\def\fnum@ptable{Table \thetable}
\def\fnum@ptablecont{Table \thetable---{\rm Continued}}

\newdimen\pt@tmpcapwidth
\def\set@mkcaption{\long\def\@makecaption##1##2{\ifdim\pt@width>\z@%
\pt@tmpcapwidth\pt@width\else\pt@tmpcapwidth\textwidth\fi%
\center\parbox{\pt@tmpcapwidth}{\center\rm##1.\quad##2\endcenter}%
\endcenter}}
\def\set@mkcaptioncont{\long\def\@makecaption##1##2{
\center\rm##1\endcenter\vskip 2.5ex}}

\newenvironment{deluxetable}[1]{\def\pt@format{\string#1}%
\set@tblnotetext\global\pt@ncol=0\global\pt@column=0\global\pt@page=1%
\let\footnotesize=\footnotesave%
\def\pt@addcol{\global\advance\pt@ncol by\@ne}}%
{%
\pt@width\wd\pt@box\box\pt@box\spew@ptblnotes%
\typeout{Table \thetable\space has been set to width \the\pt@width}%
\endcenter\end@float%
\let\footnotesize=\normalsize}

\def\startdata{\pt@line=0\pt@calcnlines%
\ifdim\pt@width>\z@\def\@halignto{to \pt@width}\else\def\@halignto{}\fi%
\let\fnum@table=\fnum@ptable\set@mkcaption%
\@float{table}\center\caption{\pt@caption}\leavevmode%
\setbox\pt@box=\pt@tabular{\pt@format}\pt@head}

\def\pt@nl{\global\advance\pt@line by\@ne%
\ifnum\pt@line=\pt@nlines%
\endtabular\pt@width\wd\pt@box\box\pt@box
\typeout{Page \the\pt@page\space of table \thetable\space has been set to
width \the\pt@width\space with \the\pt@nlines\space lines per page}%
\global\advance\pt@page by\@ne%
\endcenter\end@float\clearpage%
\addtocounter{table}{\m@ne}%
\let\fnum@table=\fnum@ptablecont\set@mkcaptioncont%
\@float{table}\center\caption{\pt@caption}\leavevmode%
\global\pt@ncol=\pt@column
\global\pt@line=0%
\setbox\pt@box=\pt@tabular{\pt@format}\pt@head%
\else\\
\fi}

\let\nl=\pt@nl
\let\nextline=\pt@nl

\def\tablebreak{\pt@line\pt@nlines\advance\pt@line by\m@ne\pt@nl}

\def\cutinhead#1{\noalign{\vskip 1.5ex}
\hline\pt@nl\noalign{\vskip -4ex}
\multicolumn{\pt@ncol}{c}{#1}\pt@nl
\noalign{\vskip .8ex}
\hline\pt@nl\noalign{\vskip -2ex}}

\def\sidehead#1{\noalign{\vskip 1.5ex}
\multicolumn{\pt@ncol}{@{\hskip\z@}l}{#1}\pt@nl
\noalign{\vskip .5ex}}

\def\set@tblnotetext{\def\tablenotetext##1##2{{%
\@temptokena={\vspace{0ex}{%
\parbox{\pt@width}{\hskip1em$^{\rm ##1}$##2}\par}}%
\@temptokenb=\expandafter{\tblnote@list}%
\xdef\tblnote@list{\the\@temptokenb\the\@temptokena}}}}

\def\spew@ptblnotes{
\ifx\tblnote@list\@empty\relax
\else
\par
\vspace{2ex}
{\parskip=1.5ex%
\tblnote@list}
\gdef\tblnote@list{}
\fi}

\def\tablerefs#1{\@temptokena={\vspace*{3ex}{%
\parbox{\pt@width}{\hskip1em\rm References. --- #1}\par}}%
:
\@temptokenb=\expandafter{\tblnote@list}
\xdef\tblnote@list{\the\@temptokenb\the\@temptokena}}

\def\tablecomments#1{\@temptokena={\vspace*{3ex}{%
\parbox{\pt@width}{\hskip1em\rm Note. --- #1}\par}}%
\@temptokenb=\expandafter{\tblnote@list}
\xdef\tblnote@list{\the\@temptokenb\the\@temptokena}}
\@ifundefined{epsfbox}{\@input{epsf.sty}}{\relax}
\def\eps@scaling{.95}
\def\epsscale#1{\gdef\eps@scaling{#1}}

\def\plotone#1{\centering \leavevmode
\epsfxsize=\eps@scaling\textwidth \epsfbox{#1}}

\newif\if@finalstyle \@finalstylefalse
\if@finalstyle
\ps@myheadings		
\let\ps@title=\ps@paspcstitle	
\else
\ps@plain			
\let\ps@title=\ps@plain	
\fi
\ds@twoside
\def\astropbibitem{\@lbibitem}

\def\@lbibitem#1#2#3{\item[\hfill]\if@filesw 
      { \def\protect##1{\string ##1\space}\immediate
        \write\@auxout{\string\astropbibcite{#3}{#1}{#2}}\fi\ignorespaces}}

\def\astropbibcite#1#2#3{\global\@namedef{b@#1}{#2\ #3} 
			\global\@namedef{newb@#1}{#2\ (#3}
			\global\@namedef{nameb@#1}{#2}
			\global\@namedef{yearb@#1}{#3}}

\let\citation\@gobble

\def\cite{\@ifnextchar [{\@tempswatrue\@citex}{\@tempswafalse\@citex[]}}

\def\citeone{\@ifnextchar [{\@tempswatrue\@newcitex}
			   {\@tempswafalse\@newcitex[]}}

\def\citename{\@ifnextchar [{\@tempswatrue\@namecitex}
			   {\@tempswafalse\@namecitex[]}}

\def\citeyear{\@ifnextchar [{\@tempswatrue\@yearcitex}
			   {\@tempswafalse\@yearcitex[]}}

\def\@citex[#1]#2{\if@filesw\immediate\write\@auxout{\string\citation{#2}}\fi
  \def\@citea{}\@cite{\@for\@citeb:=#2\do
     {\@citea\def\@citea{; }\@ifundefined
       {b@\@citeb}{{\bf ?}\@warning
       {Citation `\@citeb' on page \thepage \space undefined}}%
{\csname b@\@citeb\endcsname}}}{#1}}

\def\@newcitex[#1]#2{\if@filesw\immediate\write\@auxout{\string\citation{#2}}\fi
  \def\@newcitea{}\@newcite{\@for\@newciteb:=#2\do
     {\@newcitea\def\@newcitea{; }\@ifundefined
       {newb@\@newciteb}{{\bf ? (?}\@warning
       {Citation `\@newciteb' on page \thepage \space undefined}}%
{\csname newb@\@newciteb\endcsname}}}{#1}}

\def\@namecitex[#1]#2{\if@filesw\immediate\write\@auxout{\string\citation{#2}}\fi
  \def\@namecitea{}\@namecite{\@for\@nameciteb:=#2\do
     {\@namecitea\def\@namecitea{; }\@ifundefined
       {nameb@\@nameciteb}{{\bf ? (?}\@warning
       {Citation `\@nameciteb' on page \thepage \space undefined}}%
{\csname nameb@\@nameciteb\endcsname}}}{#1}}

\def\@yearcitex[#1]#2{\if@filesw\immediate\write\@auxout{\string\citation{#2}}\fi
  \def\@yearcitea{}\@yearcite{\@for\@yearciteb:=#2\do
     {\@yearcitea\def\@yearcitea{; }\@ifundefined
       {yearb@\@yearciteb}{{\bf ? (?}\@warning
       {Citation `\@yearciteb' on page \thepage \space undefined}}%
{\csname yearb@\@yearciteb\endcsname}}}{#1}}

\let\bibdata=\@gobble
\let\bibstyle=\@gobble

\def\bibliography#1{\if@filesw\immediate\write\@auxout{\string\bibdata{#1}}\fi
  \@input{\jobname.bbl}}

\def\bibliographystyle#1{\if@filesw\immediate\write\@auxout
    {\string\bibstyle{#1}}\fi}

\def\nocite#1{\@bsphack
  \if@filesw\immediate\write\@auxout{\string\citation{#1}}\fi
  \@esphack}

\def\@cite#1#2{({#1\if@tempswa ; #2\fi})}

\def\@newcite#1#2{{#1\if@tempswa ; #2\fi})}

\def\@namecite#1#2{#1}
\def\@yearcite#1#2{#1}

\def\thebibliography#1{\section*{References}\list
 {}{\setlength\labelwidth{1.4em}\leftmargin\labelwidth
 \setlength\parsep{0pt}\setlength\itemsep{0pt}
 \setlength{\itemindent}{-\leftmargin}
 \usecounter{enumi}}
 \def\newblock{\hskip .11em plus .33em minus -.07em}
 \sloppy
 \sfcode`\.=1000\relax}

\makeatother

\newcommand{\beq}{\begin{equation}}
\newcommand{\eeq}{\end{equation}}
\newcommand{\beqnn}{\begin{displaymath}}	
\newcommand{\eeqnn}{\end{displaymath}}		
\newcommand{\beqa}{\begin{eqnarray}}
\newcommand{\eeqa}{\end{eqnarray}}
\newcommand{\beqann}{\begin{eqnarray*}}
\newcommand{\eeqann}{\end{eqnarray*}}

\newcommand{\ben}{\begin{enumerate}}
\newcommand{\een}{\end{enumerate}}
\newcommand{\bit}{\begin{itemize}}
\newcommand{\eit}{\end{itemize}}
\newcommand{\bc}{\begin{center}}
\newcommand{\ec}{\end{center}}

\newcommand{\eqref}[1]{Equation~\ref{#1}}

\newcommand{\eBoo}{\mbox{$\eta$~Boo}}
\newcommand{\kIBoo}{\mbox{$\kappa^{\scriptscriptstyle 1}$~Boo}}
\newcommand{\kIIBoo}{\mbox{$\kappa^{\scriptscriptstyle 2}$~Boo}}

\def\Msol{\mbox{${M}_\odot$}}
\def\Lsol{\mbox{${L}_\odot$}}
\def\Rsol{\mbox{${R}_\odot$}}

\def\solar{\mbox{$_{\odot}$}}

\def\la{\mathrel{\hbox{\rlap{\hbox{\lower4pt\hbox{$\sim$}}}\hbox{$<$}}}}
\def\ga{\mathrel{\hbox{\rlap{\hbox{\lower4pt\hbox{$\sim$}}}\hbox{$>$}}}}

\def\laeq{\lower.5ex\hbox{{$\:\scriptstyle\buildrel < \over \sim\:$}}}
\def\gaeq{\lower.5ex\hbox{{$\:\scriptstyle\buildrel > \over \sim\:$}}}

\makeatletter 
 \def\sub#1{\relax\ifmmode _{\fam\z@ #1}\else
         $_{\fam\z@ #1}$\fi}
 \def\super#1{\relax\ifmmode ^{\fam\z@ #1}\else
         $^{\fam\z@ #1}$\fi}
\makeatother

\newcommand{\down}[2]{#1\sub{#2}}

\newcommand{\mycaption}[3]
{\if*#2 \caption{#3\label{#1}}
 \else  \caption[#2]{#3\label{#1}}
 \fi}

\newcommand{\comment}[1]{\relax}

\long\def\COMMENT#1\ENDCOMMENT{}
\def\ENDCOMMENT{}

\newcommand{\muHz}{\mbox{$\mu$Hz}}

\newcommand{\Teff}{\down{T}{eff}}

\begin{document}

\title{Hipparcos parallaxes for \eBoo\ and \kIIBoo: two successes for
asteroseismology}

\author{Timothy R. Bedding}
\affil{School of Physics, University of Sydney 2006, Australia}

\author{Hans Kjeldsen and J\o rgen Christensen-Dalsgaard} 

\affil{Teoretisk Astrofysik Center, Danmarks Grundforskningsfond, and
Institut for Fysik og Astronomi. Aarhus Universitet, DK-8000 Aarhus~C,
Denmark}


\keywords{oscillations; stars: individual: \eBoo, \kIIBoo}

\section{Introduction}

The release of the Hipparcos catalogue \cite{PLK97} provides an opportunity
to check results from asteroseismology.  This has already been done for the
double-mode $\delta$~Scuti star SX~Phe: \citeone{H+P97} found excellent
agreement with the parallax derived from model calculations by
\citeone{P+ChD96}.  Here we show that Hipparcos parallaxes for two other
stars are also in good agreement with oscillation results.

\section{Solar-like oscillations in \eBoo}

\eBoo\ is a bright G0 subgiant and a good target for detecting solar-like
oscillations.  We observed this star over six nights with the 2.5-m Nordic
Optical Telescope and, by monitoring equivalent widths of Balmer lines, we
detected oscillations with amplitudes at the expected level \cite{KBV95}.
We measured frequencies for thirteen individual modes in the range
750--950\,\muHz\ and determined the large frequency separation to be
\[
  \Delta\nu = 40.3 \pm 0.3\,\muHz.
\]  
The measured frequencies were subsequently shown to be consistent with
models by Christensen-Dalsgaard, Bedding \& Kjeldsen
(\citeyear{ChDBK95note}, hereafter CBK95) and also by \citeone{Gu+D96}.  In
the light of a more accurate luminosity, we can revisit these results.
Note that an attempt by \citeone{BKK97} to confirm oscillations in \eBoo{}
using Doppler measurements was not successful.  Nevertheless, for the
present we continue to assume the reality of the detection.

\begin{table}
\begin{center}
\caption[]{Parameters of \eBoo} \label{tab.eBoo}
\smallskip
\begin{tabular}{lcc}
\hline
\noalign{\smallskip}
	& Old\tablenotemark{a}	& New\\
\hline
\noalign{\smallskip}
Parallax (mas)	& $85.8\pm 2.3$\tablenotemark{b}
		& $88.17 \pm 0.75$\tablenotemark{c}\\
Effective temperature (K) &
	 \multicolumn{2}{c}{$6050 \pm 60$\tablenotemark{d}}\\
Luminosity (\Lsol)	& $9.46 \pm 0.65$\tablenotemark{e}
		& $9.02 \pm 0.22$\tablenotemark{f} \\
Radius	(\Rsol)	& $2.81 \pm 0.08$
		& $2.74 \pm 0.036$ \\
\end{tabular}
\tablenotetext{a}{as adopted by CBK95}
\tablenotetext{b}{\citeone{HDK93}}
\tablenotetext{c}{Hipparcos Main Catalogue ({\tt
	http://vizier.u-strasbg.fr/cgi-bin/VizieR})} 
\tablenotetext{d}{\citeone{B+G89} and \citeone{B+LG94}}
\tablenotetext{e}{using the above $\Teff$, plus the angular diameter of 
	$2.24 \pm 0.02$\,mas given by \citeone{B+LG94}.  }
\tablenotetext{f}{using $V = 2.68 \pm 0.01$, $BC - BC\solar = 0.03 \pm
0.01$ and $M_V\solar = 4.825 \pm 0.01$.}
\end{center}
\end{table}

The parameters of \eBoo\ are summarised in Table~\ref{tab.eBoo}.  Our
adopted luminosity in CBK95 was based on the best available parallax.  The
more precise Hipparcos parallax, while being consistent with the
ground-based value, implies a slightly lower luminosity.  Also note that in
CBK95 we calculated the luminosity using published estimates of the
effective temperature and angular diameter, which were based on the
infrared flux method.  The calculation, also used by \citeone{Gu+D96}, was
indirect and here we prefer to use $V$-band photometry directly, as
explained in the Table.  The new luminosity is accurate to 2.4\%, an
improvement by a factor of three over the value adopted in CBK95.

\begin{figure}[p]
\vspace*{2cm}
\centerline{
\psfig{%
bbllx=73pt,bblly=93pt,bburx=512pt,bbury=677pt,%
angle=90,%
figure=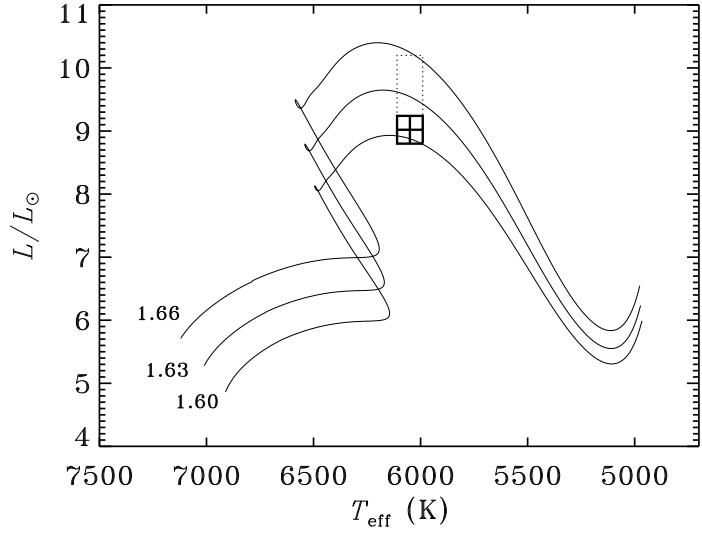,width=9.2cm}}

\caption[]{\label{fig.eBoo1} Evolutionary tracks in the H-R diagram for
three masses (labelled in \Msol).  The error boxes show $(\Teff,L)$ for
\eBoo\ adopted by CBK95 (dotted lines) and the new values (bold lines).

}
\end{figure}

\begin{figure}[p]
\centerline{
\psfig{%
bbllx=43pt,bblly=119pt,bburx=501pt,bbury=683pt,%
angle=90,%
figure=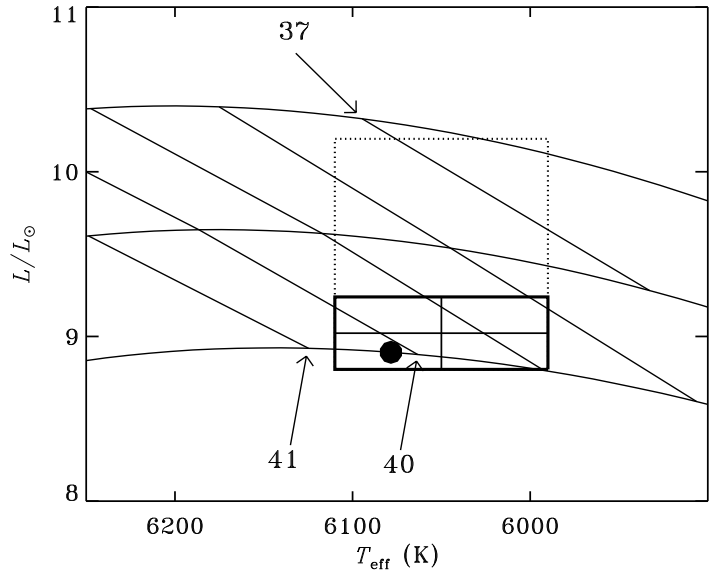,width=9.2cm}}

\caption[]{\label{fig.eBoo2} Close-up of the H-R diagram in the vicinity of
{\eBoo}.  Same as Fig.~\ref{fig.eBoo1}, with the addition of diagonal lines
which join models of constant $\Delta\nu$ (37, 38, \ldots, 41\muHz),
labelled in a few cases by $\Delta\nu$ (in $\muHz$).  The solid point
indicates the model chosen by CBK95. }

\end{figure} 

This improved luminosity constrains the expected oscillation frequencies
for \eBoo.  In Figures~\ref{fig.eBoo1} and~\ref{fig.eBoo2} we show the
location of \eBoo\ in the H-R diagram (these are similar to Figs.~1 and~2
of CBK95).  The evolution tracks use the known metallicity of \eBoo\ ($Z =
0.03$) and the solar value for the ratio of the mixing length to the
pressure scale height.  Full details of the calculations are given in
CBK95.

The diagonal lines in Figure~\ref{fig.eBoo2} join models of constant
$\Delta\nu$.  The solid point indicates the 1.6-\Msol\ model chosen by
CBK95 to have the frequency separation observed by \citeone{KBV95}.  {\em
It is clear that the improved luminosity for \eBoo\ is in excellent
agreement with the observed frequency separation.}

\citeone{Gu+D96} also computed models for \eBoo.  By matching the observed
oscillation frequencies, they derived a mass of $1.55 \pm 0.03\,\Msol$ and
predicted a parallax of $89.5 \pm 0.5$\,mas.  Their parallax agrees with
the Hipparcos measurement, again giving strong support to the reality and
interpretation of the oscillation signal.



\section{The $\delta$~Scuti star \kIIBoo}

The Aarhus group has also studied the binary system consisting of \kIBoo\
(HR~5328; $V=6.69$; F1\,V) and \kIIBoo\ (HR~5329; $V=4.54$; A8\,IV).  The
brighter component \kIIBoo\ is a $\delta$~Scuti variable.  Based on a model
fit to four observed frequencies, \citeone{FJK95} derived a distance of
47.9\,pc.


The Hipparcos parallax for this system is $21.03 \pm 0.83$\,mas, which
implies a distance of $47.6 \pm 1.9$\,pc.  This is in excellent agreement
with the distance derived by \citename{FJK95}.  However, we note that their
frequency identifications did not allow for rotational splitting, despite
the fact that \kIIBoo\ is known to be a rapid rotator.  Unless the accuracy
of the predicted parallax is coincidental, we appear to have confirmed
their assumption that the observed modes have $m=0$.

\section{Conclusion}

The results presented here for \eBoo\ and \kIIBoo, together with those for
SX~Phe by \citeone{H+P97}, all rely on the same stellar evolution
calculations \cite{ChD82}.  The fact that asteroseismic analysis has been
successfully performed for three stars covering a range of masses and
evolutionary states is an important validation of the models.

\acknowledgments

We are indepted to the Hipparcos group for making their wonderful catalogue
available on the Web.  This work was supported by the Australian
Research Council, and by the Danish National Research Foundation through
its establishment of the Theoretical Astrophysics Center.

\end{document}